\documentclass[a4paper,11pt]{article}

\usepackage{a4}
\usepackage{psfig}
\usepackage{feynmf}

\newcommand{\erf}{{\rm erf}}

\title{A theoretical model of neuronal population coding of stimuli with both continuous and discrete dimensions}

\author{Valeria Del Prete\thanks{delprete@sissa.it}  and Alessandro Treves \\
SISSA, Programme in Neuroscience, via Beirut 4, Trieste, Italy}

\begin{document}

\maketitle

\bibliographystyle{unsrt}
\begin{abstract}
In a recent study the initial rise of the mutual information between the firing rates of $N$ neurons and a set of $p$ discrete stimuli 
has been analytically evaluated, under the assumption that neurons fire independently of one another 
to each stimulus and that each conditional distribution of firing rates is gaussian.
Yet real stimuli or behavioural correlates are high-dimensional, with both discrete and continuously varying features.
Moreover, the gaussian approximation implies negative firing rates, which is biologically implausible.
Here, we generalize the analysis to the case where the stimulus or behavioural correlate
has both a discrete and a continuous dimension, like orientation and shape could be in a visual stimulus, or type and direction
in a motor action. 
The functional relationship between the firing patterns and the continuous correlate is expressed through the tuning curve of the neuron, using two different parameters to  modulate its width and its flatness.
In the case of large noise we evaluate the mutual information up to the quadratic 
approximation as a function of population size.
We also show  that in the limit of large $N$ and assuming that neurons can discriminate between continuous values with a resolution 
$\Delta \vartheta$, the mutual information grows to infinity like $log(1/\Delta \vartheta)$ when $\Delta \vartheta$ goes to zero.
Then we consider a more realistic distribution of firing rates, truncated at zero, and we prove that the resulting correction,
with respect to the gaussian firing rates, can be expressed simply as a renormalization of the noise parameter.
Finally, we demonstrate the effect of averaging the distribution 
across the discrete dimension, evaluating the mutual information only with respect to the
continuously varying correlate.

\end{abstract}

\section*{Introduction}
The strategy used by populations of neurons to code for external stimuli or behavioural correlates is a major issue which has been 
recently investigated both through data analysis and theoretical modelling. The mutual information between external correlates and 
the spiking activity of the population is one way to assess such coding quantitatively \cite{Bia+91}. Several analyses have focused on 
the coding of a {\em discrete} set of stimuli (\cite{pap35}, see \cite{proc30} for a review), which is the paradigm used in 
many experiments\cite{Opt+87,Opt+90,pap21,Rol+97a}. In this situation the mutual information is bounded by the entropy of the
stimulus set. Some theoretical studies have also considered the coding of stimuli varying in a {\em continuous} domain
\cite{seu+93,sompo+00}, which is interesting with respect to basic properties like orientation in visual stimuli, frequency in 
auditory stimuli, velocity and position in motor actions. 
In particular in \cite{sompo+00} the authors have studied the asymptotic
(large population) behaviour of the mutual information, with respect to a stimulus with a continuously varying dimension.

Yet no study has been proposed so far considering a mixture of both continuous and discrete features, which is obviously closer to 
real world stimuli or behavioural correlates.
Moreover the initial rise of the mutual information for small but increasing population size is more relevant for a comparison with 
estimates from real data, at least as far as the possibility of having simultaneous recordings from very large populations of neurons 
is restricted to very few cases. 

We have recently analyzed data recorded in the motor areas of behaving monkeys, in the laboratory of Eilon Vaadia. The monkeys
moved a manipulandum in several possible directions (approximating a continuous correlate) and with different combinations of
arms (4 {\em types} of movement, i.e. a discrete correlate). In trying to characterize the neural coding of these movements,
we were particularly interested in whether, as it is reasonable to expect, different motor areas differ, at least quantitatively,
in their coding properties. The results, obtained from records of activity in areas M1 and SMA, will be reported elsewhere
\cite{vale}; but they suggest the importance of developing theoretical models of how populations of neurons might code
simultaneously discrete and continuous correlates. For example, one clear conclusion has been that type and direction are 
not {\em independent} dimensions of the movement, in the sense for example that the information about direction, extracted from the
activity recorded with all movement types, is much lower (roughly half) of the average information about direction, obtained
with a single movement type. 

Can we embody similar properties in a model of the scheme used by neurons to code movements?
What would then be the dependence of the mutual information on the number of the possible types of movement?
How would it depend on the resolution with which the continuous correlate is sampled? 
How on the level of noise affecting the firing patterns?

In a recent work \cite{pap35} some of these questions have been investigated for a set of $p$ discrete stimuli, under 
the assumption that neurons fire independently of one another to each stimulus and that the distribution of the firing rates 
is gaussian. The linear and the quadratic approximations to the mutual information as a function of population size
were studied analytically, in the limit of large noise, as well as the approach to the ceiling in the case of small noise. 
We generalize this study considering both a discrete and a continuous dimension in the stimulus, referring specifically  
to motor actions characterized by a direction and a ``type''. Nonetheless, our model is equally applicable to other complex correlates.
We introduce a more realistic conditional firing rate distribution, than the simple gaussian one, and find a simple resulting 
correction to the gaussian model: the analytical expression of the mutual information remains the same,
except for a renormalization of the expansion parameter.
 
We then evaluate the information loss when the original activity distribution is averaged across the discrete correlate,
as is sometimes the case in the analysis of real data, and the 
mutual information is evaluated solely with respect the continuously varying feature.
Averaging out dimensions in the stimulus corresponds to losing accuracy in its 
description, and hence the information loss.

Our theoretical analysis allows a direct comparison with real curves; we present one 
comparison and discuss possible causes for the discrepancy between data and model. 
In particular, correlations between neurons, that are not included in the model, 
might play a relevant role, enhancing or 
decreasing redundancy in population coding \cite{pan+99}. 
This issue will be the object of future work.   

\section*{The gaussian model} 
First, we consider a coding scheme where the distribution of the firing rates 
conditional to the movements is gaussian, similarly to the case examined in  
\cite{pap35}. This assumption implies that negative rates have a non zero 
probability to occur, but it allows an easier analytical treatment.
We will examine  a more realistic scheme later on. 

Consider the following distribution:

\begin{equation}
P(\{\eta_i\}|\vartheta,s)=\prod_{i=1}^N \frac{1}{\sqrt{2\pi\sigma^2}}
exp-\left[\left(\eta_i-\tilde{\eta}_i(\vartheta,s)
\right)^2/2\sigma^2\right];
\label{dist}
\end{equation}
$\eta_i$ is the firing rate of neuron $i$; $\vartheta$ and $s$ parameterize
respectively the direction and the type of movement; $\tilde{\eta}_i(\vartheta,s)$ is the 
average firing rate of the neuron with the movement parameterized by $\vartheta$,$s$.

\begin{figure}
\centerline{
\psfig{figure=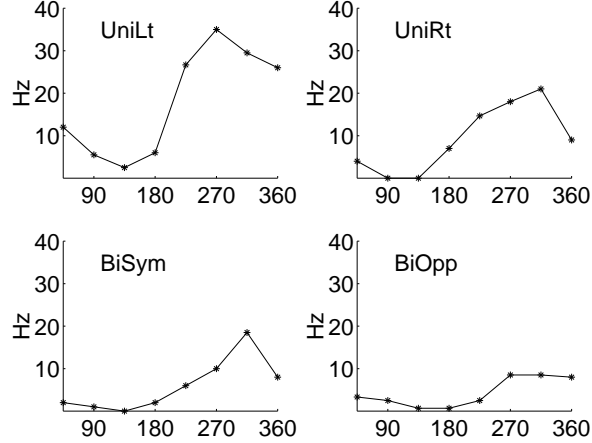,height=6cm,angle=0}}
\caption{Directional tuning for a cell recorded in the right supplementary motor area of a monkey performing 4 different types 
of arm movement. UniLt=unimanual left; UniRt=unimanual right; BiSym=bimanual symmetric; BiOpp=bimanual opposite. 
Notice that the type of movement strongly modulates the amplitude of directional tuning, but not so much its preferred direction.}
\label{tuning_data}
\end{figure}

In general, the directional tuning of real cells in motor cortices is modulated by the type of movement performed. 
We show an example of this modulation, with the typical shape of tuning curves, in fig.\ref{tuning_data} (data kindly provided by Eilon Vaadia). The modulation of the preferred direction looks weaker than the overall amplitude modulation.

\begin{figure}
\hbox{
\mbox{(a)}
\psfig{figure=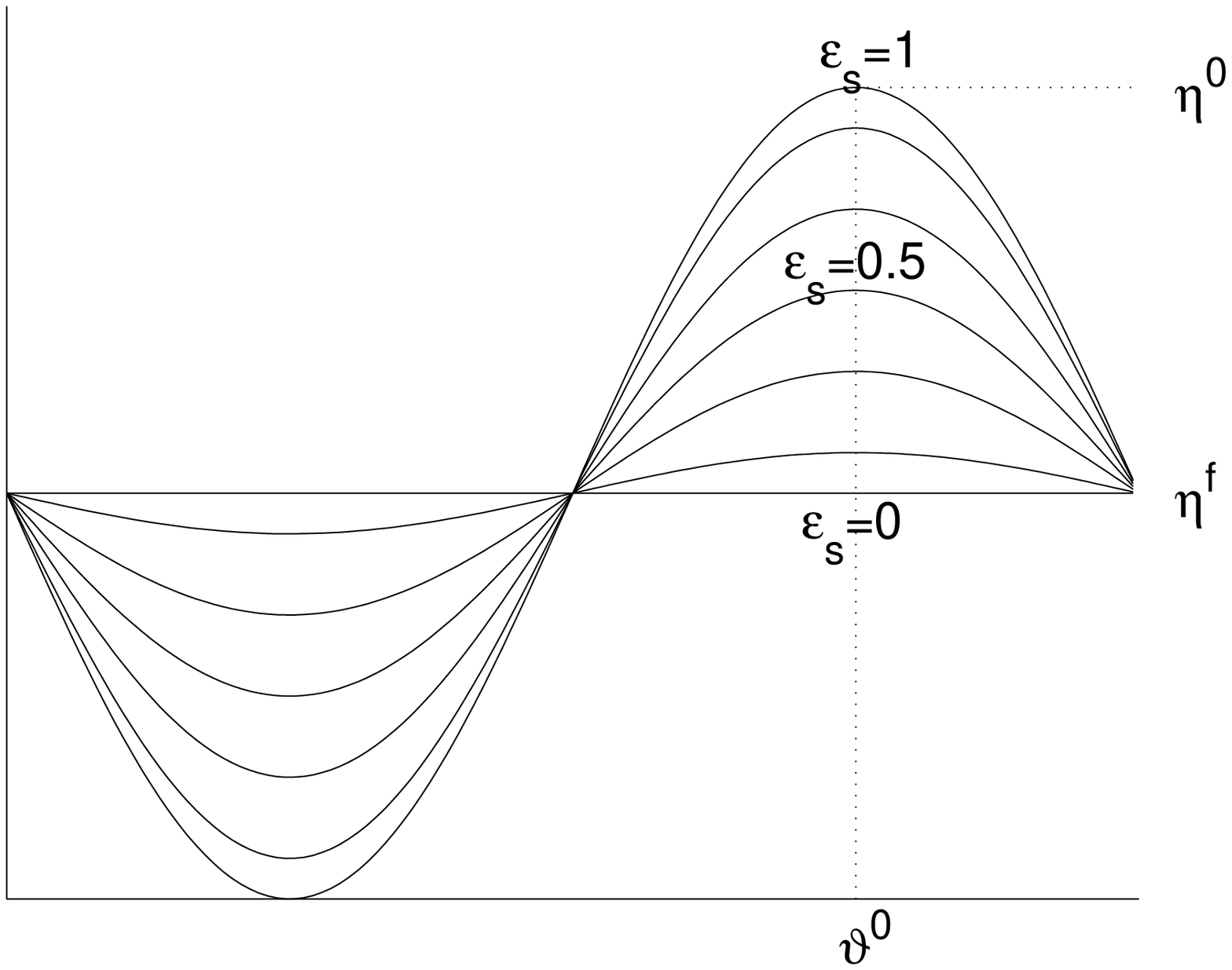,height=6cm,angle=0}
\mbox{(b)}
\psfig{figure=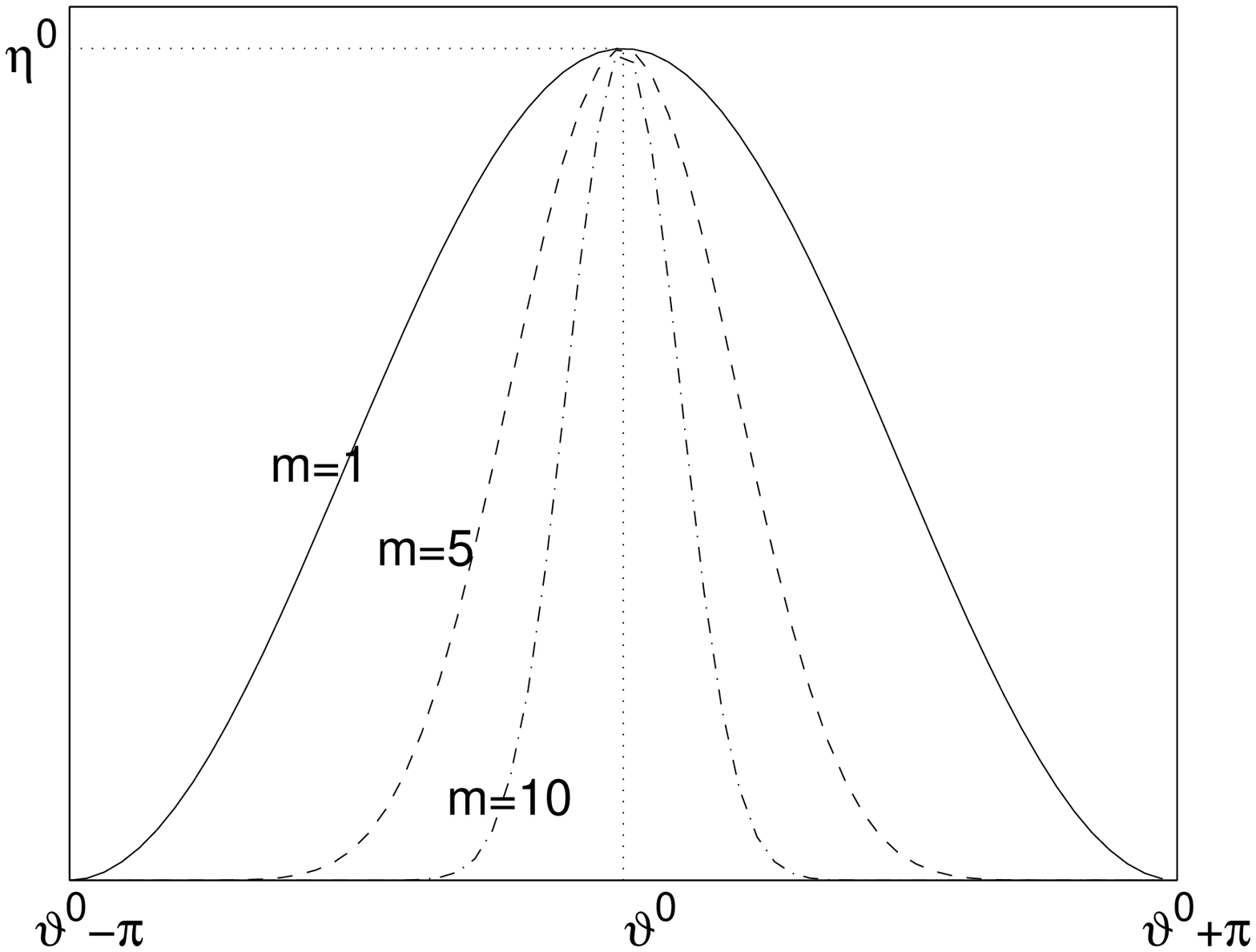,height=6cm,angle=0}}
\caption{Cosinusoidal tuning curves as in eqs.(\protect\ref{tuning_tot}), (\protect\ref{tuning2}). (a) $m=1$; modulation for different values of $\varepsilon_s$. (b) $\varepsilon_s=1$; modulation for different values of $m$.}
\label{tuning_2}
\end{figure}

For our model we have considered the following function:

\begin{equation}
\tilde{\eta}_i(\vartheta,s)=\varepsilon^i_s\bar{\eta}_i(\vartheta)-(1-\varepsilon_s^i)\eta_i^f;
\label{tuning_tot}
\end{equation}

\begin{equation}
\bar{\eta}_i(\vartheta-\vartheta^0_i)=\eta^0_i \cos^{2m}\left(\frac{\vartheta-\vartheta^0_i}{2}\right)=\eta^0_i\left[\frac{1}{2^{2m}}\left(\begin{array}{c}2m\\m\end{array}\right)+\frac{1}{2^{2m-1}}\sum_{\nu=0}^{m-1}\left(\begin{array}{c}2m\\\nu\end{array}\right)\cos\left\{\left(m-\nu\right)\left(\vartheta-\vartheta^0_i\right)\right\}\right];
\label{tuning2}
\end{equation}
$\varepsilon_s^i$ is a quenched random variable distributed between $0$
and $1$. 
The meaning of eq.(\ref{dist}),(\ref{tuning_tot}),(\ref{tuning2}) is that the firing rate of neuron $i$ given the movement parameterized by ($\vartheta$,$s$) follows a gaussian distribution 
centered around a tuning curve 
$\bar{\eta}_i(\vartheta)$ whose flatness is modulated through the parameters 
$\varepsilon_s^i$. If $\varepsilon^i_s$ is zero for some particular $s$ 
the firing of the cell does not depend, for that movement type, on the direction of the movement. On the other hand 
if $\varepsilon_s^i$ assumes a fixed value for all $s$ the directional tuning does not modulate with the type of movement.
Tuning curves with a cosinusoidal shape have already been considered 
to model the directional selectivity of sensory neurons \cite{seu+93}. 
Fig.\ref{tuning_2} shows the amplitude of the tuning curve.

We neglect the modulation of the preferred direction with the type, as it would burden the analytical calculations; 
it will be the object of future analyses. 

\section*{Evaluation of the mutual information for the gaussian model}

We are interested in the mutual information \cite{sha+48} between the neuronal firing 
rates and the movements:

\begin{equation}
I(\{\eta_i\},\vartheta\otimes s)=\left\langle\sum_{s=1}^p\int \!d\vartheta\! \int \prod_i d\eta_iP(\vartheta,s) P(\{\eta_i\}|\vartheta,s)\log_2\frac{P(\{\eta_i\}|\vartheta,s)}{P(\{\eta_i\})}\right\rangle_{\varepsilon,\vartheta^0};
\label{info}
\end{equation}

where the distribution $P(\eta_i|\vartheta,s)$ is given in eq.(\ref{dist}) and  
$\langle .. \rangle_{\varepsilon,\vartheta^0}$ is a short notation for  the average 
across the quenched variables
$\{\varepsilon_s^i\}$,$\{\vartheta_i^0\}$.
In fact we are not interested in a particular realization of the tuning, but in the 
average across all its possible realizations.
 
Eq.(\ref{info}) can be written as:

\begin{equation}
I(\{\eta_i\},\vartheta\otimes s)=H(\{\eta_i\})-\left\langle H(\{\eta_i\}|\vartheta,s)
\right\rangle_{\vartheta,s};
\end{equation}

\begin{equation}
\left\langle H(\{\eta_i\}|\vartheta,s)\right\rangle_{\vartheta,s}=
\left\langle\sum_{s=1}^p\int \!d\vartheta\! \int \prod_i d\eta_i 
P(\vartheta,s) P(\{\eta_i\}|\vartheta,s)\log_2
P(\{\eta_i\}|\vartheta,s)
\right\rangle_{\varepsilon,\vartheta^0};
\label{equiv}
\end{equation}

\begin{equation}
H(\{\eta_i\})=\left\langle\sum_{s=1}^p\int \!d\vartheta\! \int \prod_i d\eta_i 
P(\vartheta,s) P(\{\eta_i\}|\vartheta,s)\log_2\left[
\sum_{s^\prime=1}^p\int \!d\vartheta^\prime P(s^\prime,\vartheta^\prime)
P(\{\eta_i\}|\vartheta^\prime,s^\prime)\right]
\right\rangle_{\varepsilon,\vartheta^0};
\label{outent}
\end{equation}

The calculation of the equivocation $\left\langle H(\{\eta_i\}|\vartheta,s)\right\rangle_{\vartheta,s}$
is straightforward and the result is additive in the population size:

\begin{equation}
\left\langle H(\{\eta_i\}|\vartheta,s)\right\rangle_{\vartheta,s}=
\frac{N}{2\ln2}[1+\ln(2\pi\sigma^2)];
\label{final_eq}
\end{equation}

The linearity in $N$  is standard whenever the conditional distribution of firing 
rates can be factorized across neurons, as in  
eq.(\ref{dist}). 

The evaluation of the rate entropy $H(\{\eta_i\})$ can be carried out introducing  $n$ 
replicas \cite{meza+87,treves+95} 
for both the discrete and the continuous dimensions $\vartheta$ and $s$, which allows
to get rid of the logarithm in eq.(\ref{outent}): 

\begin{eqnarray}
&&H(\{\eta_i\})=-\frac{1}{\ln2}\lim_{n\rightarrow 0}\frac{1}{n}
\left (\sum_{s_1..s_{n+1}=1}^p\int \!d\vartheta_1..d\vartheta_{n+1}
\int\prod_i d\eta_i \frac{1}{(2\pi p)^{n+1}}\right.\nonumber\\
&&\mbox{x}\left. \left\langle \prod_{i=1}^N\prod_{k=1}^{n+1}
\frac{1}{\sqrt{2\pi\sigma^2}}
exp-\left[\left(\eta_i-\varepsilon^i_{s_k}\bar{\eta}_i(\vartheta_k)-(1-\varepsilon_{s_k}^i)
\eta_i^f\right)^2/2\sigma^2\right]\right\rangle_{\varepsilon,\vartheta^0}
-1\right );
\end{eqnarray}

Integrating over $\{\eta_i\}$ and rearranging terms one obtains:

\begin{equation}
H(\{\eta_i\})=-\frac{1}{\ln2}\lim_{n\rightarrow 0}\frac{1}{n}
\left (\sum_{s_1..s_{n+1}=1}^{p}\int \!d\vartheta_1..d\vartheta_{n+1}
\frac{(n+1)^{-\frac{N}{2}}}{(2\pi p)^{n+1}}\frac{1}{(\sqrt{2\pi\sigma^2})^{nN}}
\left\langle \prod_{i=1}^N exp(-R_i)\right\rangle_{\varepsilon,\vartheta^0}
-1\right );
\label{entropy}
\end{equation}

\begin{equation}
R_i=\sum_{k,l}\left (
\varepsilon_{s_k}^i(\bar{\eta}_i(\vartheta_k)-\eta_i^f)-
\varepsilon_{s_l}^i(\bar{\eta}_i(\vartheta_l)-\eta_i^f)\right )^2/
(4\sigma^2 (n+1));
\label{gen_R}
\end{equation}

\subsection*{Large $\boldmath{\mbox{$\sigma$}}$ limit}

An exact analytical evaluation of eq.(\ref{entropy}) is not possible without resorting to some approximation. 
In line with the analysis performed in \cite{pap35} we assume now that the quenched randomness is uncorrelated and identically 
distributed across neurons:

$$
\varrho(\{\varepsilon^i_s\})=\left(\prod_s\varrho(\varepsilon_s)\right)^N;
$$
$$
\varrho(\{\vartheta^0_i\})=\left[\varrho(\vartheta^0)\right]^N=\frac{1}{(2\pi)^N};
$$

We assume also that $\eta^f_i=\eta^f$ $\forall i$.
Then one can write:

$$
\left\langle \prod_i exp(- R_i)\right\rangle_{\varepsilon,\vartheta^0}=
\left\langle exp(-R)\right\rangle^N_{\varepsilon,\vartheta^0};
$$

We consider now the limit of large noise $\sigma$; in this case, since $R\simeq 1/\sigma^2$  
we can expand $exp(-R)$; keeping only terms of order $(N/\sigma^k)^l$, with $k\leq 2$ and $l=1,2$
we obtain:

\begin{eqnarray}
&&\left\langle exp(-R)\right\rangle^N_{\varepsilon,\vartheta^0}\simeq 1-N\left\langle R\right\rangle_{\varepsilon,\vartheta^0}+\frac{N^2}{2}\left\langle R\right\rangle^2_{\varepsilon,\vartheta^0}=\nonumber\\
&& 1-\frac{N}{4\sigma^2(n+1)}\left\langle \sum_{k,l\neq k}\left (
\varepsilon_{s_k}(\bar{\eta}(\vartheta_k)-\eta^f)-
\varepsilon_{s_l}(\bar{\eta}(\vartheta_l)-\eta^f)\right )^2\right\rangle_{\varepsilon,
\vartheta^0}\!\!+\frac{1}{2}\frac{N^2}{(4\sigma^2(n+1))^2}\nonumber\\
&&\sum_{k,l\neq k}\sum_{\varrho,\mu\neq \varrho}\!
\left\langle\left (\varepsilon_{s_k}(\bar{\eta}(\vartheta_k)\!-\!\eta^f)\!-\!
\varepsilon_{s_l}(\bar{\eta}(\vartheta_l)\!-\!\eta^f)\right )^2\right\rangle_{\varepsilon,\vartheta^ 0}\left\langle
\left (\varepsilon_{s_\varrho}(\bar{\eta}(\vartheta_\varrho)\!-\!\eta^f)\!-\!
\varepsilon_{s_\mu}(\bar{\eta}(\vartheta_\mu)\!-\!\eta^f)\right )^2
\right\rangle_{\varepsilon,\vartheta^0}
\label{R_appr}
\end{eqnarray}

To first order in $N/\sigma^2$ we obtain:

\begin{eqnarray}
&&H(\{\eta_i\})=-\frac{1}{\ln2}\lim_{n\rightarrow 0}\frac{1}{n}
\left(\sum_{s_1..s_{n+1}=1}^p\int \!d\vartheta_1..d\vartheta_{n+1}
\frac{(n+1)^{-\frac{N}{2}}}{(2\pi p)^{n+1}}\frac{1}{(\sqrt{2\pi\sigma^2})^{nN}}\right.\nonumber\\
&&\left.\left( 1-\frac{N}{4\sigma^2(n+1)}\left\langle\sum_{k,l\neq k}\left(
\varepsilon_{s_k}(\bar{\eta}(\vartheta_k)-\eta^f)-
\varepsilon_{s_l}(\bar{\eta}(\vartheta_l)-\eta^f)\right)^2\right\rangle_{\varepsilon,
\vartheta^0}\right )-1\right);
\end{eqnarray}

This result is valid for a generic directional tuning curve $\bar{\eta}(\vartheta)$.
We consider now our specific choice,  eq.(\ref{tuning2}), examining  the simplest case $m=1$ 
first. After averaging across direction selectivities $\{\vartheta^0\}$ and 
integrating over continuous replicas  $\vartheta_1..\vartheta_{n+1}$ we obtain

\begin{eqnarray*}
&&H(\{\eta_i\})=-\frac{1}{\ln2}\lim_{n\rightarrow 0}\frac{1}{n}
\left (\sum_{s_1..s_{n+1}=1}^p
\frac{(n+1)^{-\frac{N}{2}}}{(p)^{n+1}}\frac{1}{(\sqrt{2\pi\sigma^2})^{nN}}\right .\nonumber\\
&&\mbox{x}\left.\left(  1-\frac{N(\eta^0)^2}{4\sigma^2(n+1)}\sum_{k,l\neq k}\left [\left(\alpha(\alpha-1)+\frac{1}{4}\right)\left\langle(\varepsilon_{s_k}-\varepsilon_{s_l})^2\right\rangle_{\varepsilon}+\frac{1}{8}\left\langle\varepsilon_{s_k}^2+\varepsilon_{s_l}^2\right\rangle_{\varepsilon}\right ]\right)-1\right);
\end{eqnarray*}
where we have defined $\alpha=\eta^f/\eta^0$.
To perform the average across the quenched variables $\{\varepsilon_s\}$ we assume that they are equally 
distributed across $p$ movement types, namely:

$$
\varrho(\{\varepsilon_s\})=\prod_s\varrho(\varepsilon_s)=[\varrho(\varepsilon)]^p;
$$

In this case the sum over indexes $k$ and $l$ generates $n(n+1)$ identical terms. The summation on discrete replicas 
yields a factor $p^{n+1}$ multiplying the term $\left\langle (\varepsilon_{s_k})^2\right\rangle_{\varepsilon}$ and a factor $p^n(p-1)$
multiplying the term $\left\langle (\varepsilon_{s_k}-\varepsilon_{s_l})^2\right\rangle_{\varepsilon}$, since this last term is non zero only when $s_k\neq s_l$.

Taking the limit $n\rightarrow 0$ yields

\begin{eqnarray}
&&H(\{\eta_i\})\simeq \frac{N}{2\ln2}[1+\ln(2\pi\sigma^2)]\nonumber\\
&&+\frac{1}{\ln2}\frac{N(\eta^0)^2}{4\sigma^2}
\left[\frac{p-1}{p}2\left(\alpha(\alpha-1)+\frac{1}{4}\right)\lambda_1+\frac{1}{4}\lambda_2\right];
\label{final_ent}
\end{eqnarray}

\begin{equation}
\lambda_1=\int d\varepsilon \varrho(\varepsilon)
\varepsilon^2-\left[\int d\varepsilon\varrho(\varepsilon)\varepsilon\right]^2;
\label{lambda1}
\end{equation}

\begin{equation}
\lambda_2=\int d\varepsilon \varrho(\varepsilon) \varepsilon^2;
\label{lambda2}
\end{equation}

From eqs.(\ref{final_ent}) and (\ref{final_eq}) the final expression for the mutual information can be written
\begin{equation}
I(\{\eta_i\},\vartheta\otimes s)\simeq\frac{1}{\ln2}\frac{N(\eta^0)^2}{4\sigma^2}
\left[\frac{p-1}{p}2\left(\alpha(\alpha-1)+\frac{1}{4}\right)\lambda_1+\frac{1}{4}\lambda_2\right];
\label{final_info}
\end{equation}

In the more general case of a power $2m$ of the cosine, in eq.(\ref{tuning2}), it is easy 
to show that the final result can be expressed as 

\begin{equation}
I(\{\eta_i\},\vartheta\otimes s)\simeq\frac{1}{\ln2}\frac{N(\eta^0)^2}{4\sigma^2}
\left[\frac{p-1}{p}2\left(\alpha-A_1\right)^2\lambda_1+2\left(A_2-(A_1)^2\right)\lambda_2\right];
\label{final_info2}
\end{equation}
where:

\begin{equation}
\eta^0 A_1=\left\langle\bar{\eta}(\vartheta_k)\right\rangle_{\vartheta^0}=\eta^0\frac{1}{2^{2m}}\left(\begin{array}{c}2m\\m\end{array}\right);
\label{A_1}
\end{equation}

\begin{equation}
\left(\eta^0\right)^2A_2=\left\langle\bar{\eta}^2(\vartheta_k)\right\rangle_{\vartheta^0}=\left(\eta^0\right)^2\frac{1}{2^{4m}}\left(\begin{array}{c}4m\\2m\end{array}\right);
\label{A_2}
\end{equation}

The calculation of the coefficient of the second order term, which multiplies 
$N^2/\sigma^4$ in eq.(\ref{R_appr}), is slightly more complex and implies integration 
of terms with 4-replica interaction. The detailed analytical evaluation is given in 
the appendix. 
The final result up to the quadratic approximation reads:

\begin{eqnarray}
&&I(\{\eta_i\},\vartheta\otimes s)\simeq\frac{1}{\ln2}\left\{\frac{N(\eta^0)^2}{4\sigma^2}
\left[\frac{p-1}{p}2\left(\alpha-A_1\right)^2\lambda_1+2\left(A_2-(A_1)^2\right)\lambda_2\right]-\frac{N^2(\eta^0)^4}{2(4\sigma^2)^2}\right.\\
&&\left.\left\{\frac{p-1}{p^2}2\left(2\left(\alpha-A_1\right)^2\lambda_1\right)^2+\left[\frac{p-1}{p}\left(\lambda_1-\lambda_2\right)^2+\frac{(\lambda_2)^2}{p}\right]\left[\left(\frac{1}{2^{2m-1}}\right)^4\sum_{\nu=0}^{m-1}\left[\left(\begin{array}{c}2m\\\nu\end{array}\right)\right]^4\right]\right\}\right\};\nonumber
\label{final_info4}
\end{eqnarray}

where the expressions of $\lambda_1$,$\lambda_2$,$A_1$,$A_2$ are given respectively in eq.(\ref{lambda1}),(\ref{lambda2}),(\ref{A_1}),(\ref{A_2})
In the limit of large $p$ we have

\begin{eqnarray}
&&I(\{\eta_i\},\vartheta\otimes s)\simeq\frac{1}{\ln2}\left\{\frac{N(\eta^0)^2}{4\sigma^2}
\left[2\left(\alpha-A_1\right)^2\lambda_1+2\left(A_2-(A_1)^2\right)\lambda_2\right]\right .\nonumber\\
&&\left .-\frac{N^2(\eta^0)^4}{2(4\sigma^2)^2}\left\{\left(\lambda_1-\lambda_2\right)^2\left[\left(\frac{1}{2^{2m-1}}\right)^4\sum_{\nu=0}^{m-1}\left[\left(\begin{array}{c}2m\\\nu\end{array}\right)\right]^4\right]\right\}\right\};
\label{largep}
\end{eqnarray}

\begin{figure}
\centerline{
\psfig{figure=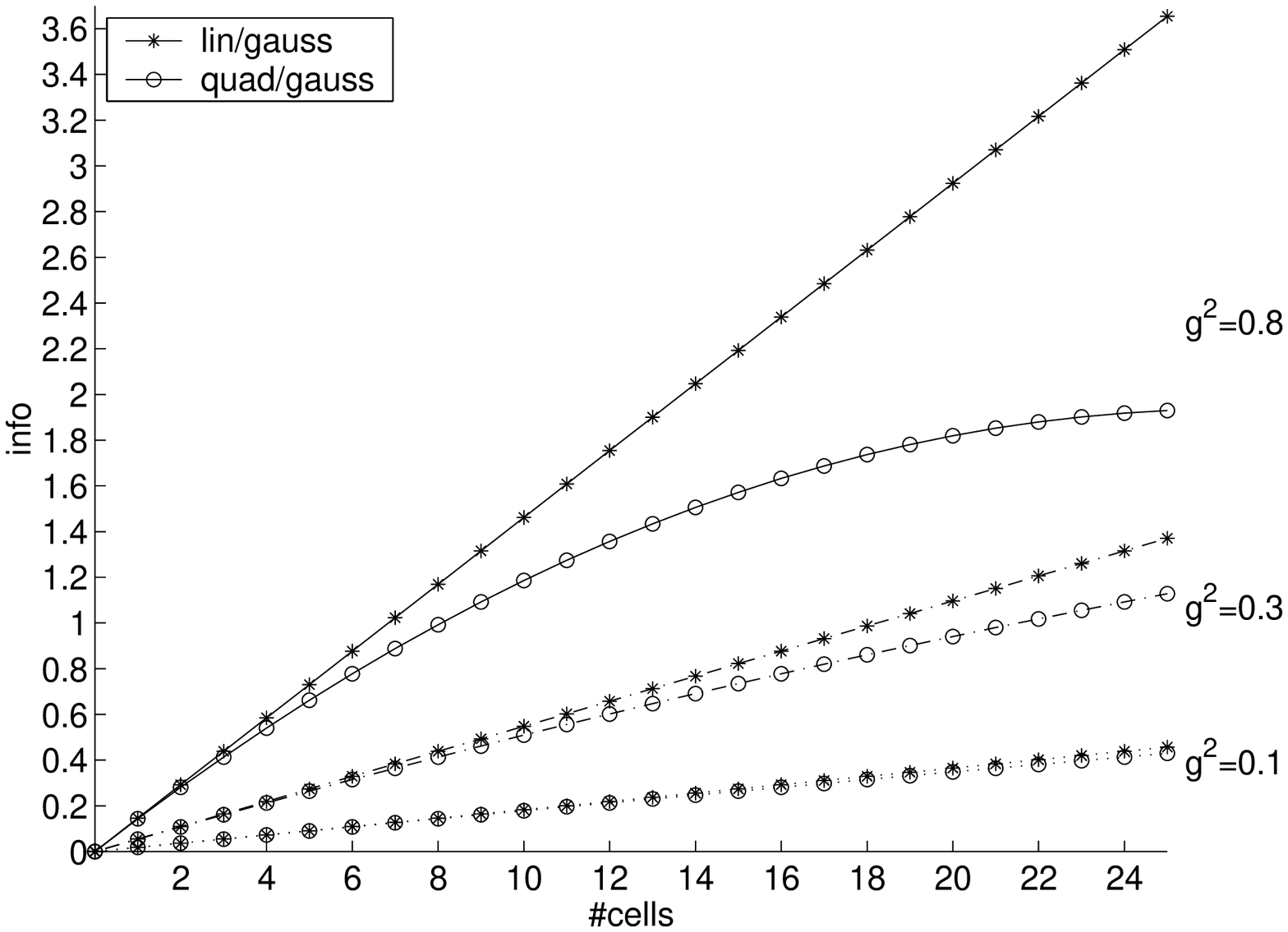,height=6cm,angle=0}}
\caption{Information rise, from eq.(\protect\ref{final_info4}), for different values of the expansion parameter 
$g^2=\left(\eta^0/2\sigma\right)^2$; $m=1$; $p=4$; the distribution $\varrho(\varepsilon)$ in eqs.(\protect\ref{lambda1}), 
(\protect\ref{lambda2}) is just equal to $1/3$ for each of the $3$ allowed $\varepsilon$ values of $0$,$1/2$ and $1$.}
\label{res_vary_g}
\end{figure}

\begin{figure}
\hbox{
\mbox{(a)}
\psfig{figure=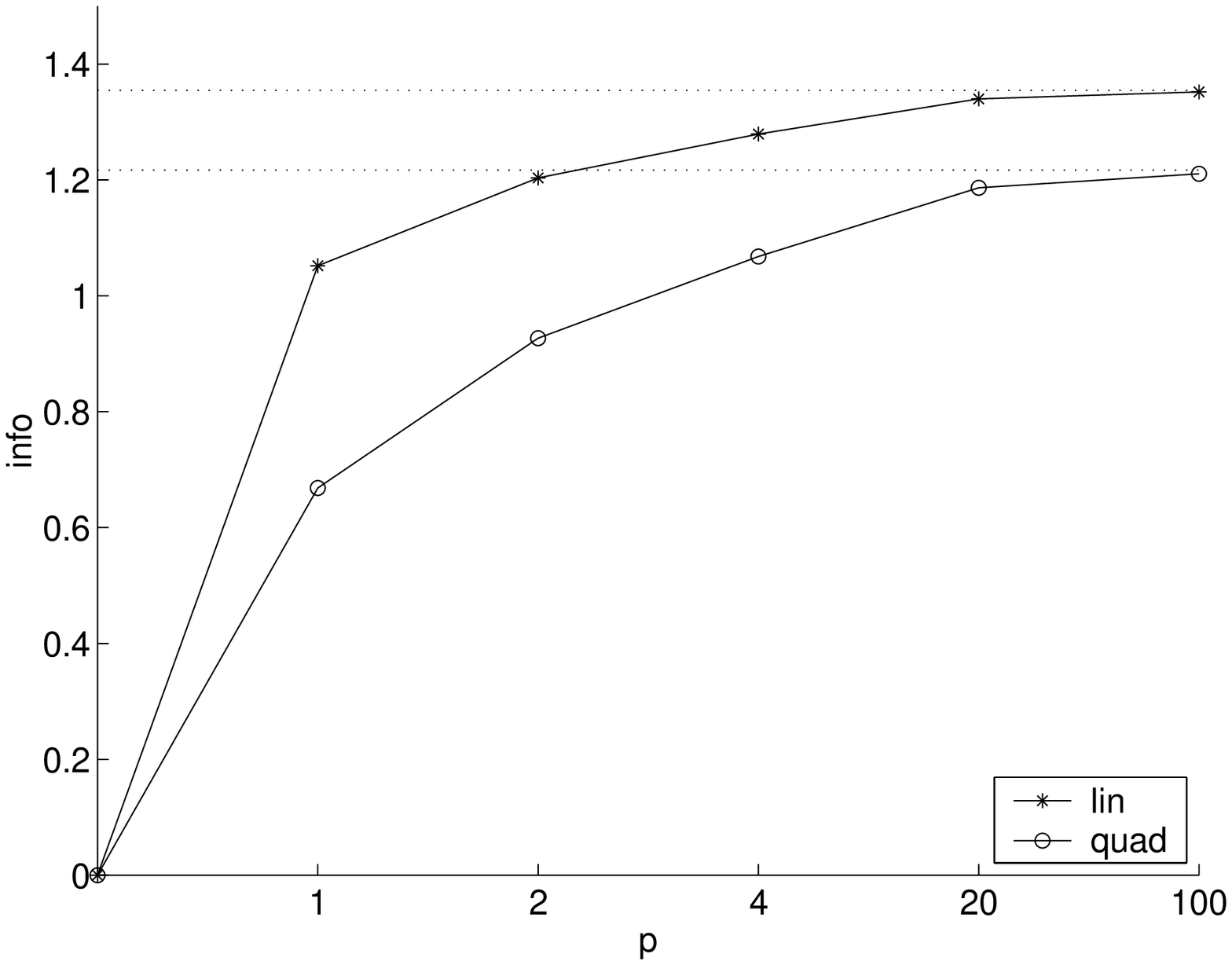,height=6cm,angle=0}
\mbox{(b)}
\psfig{figure=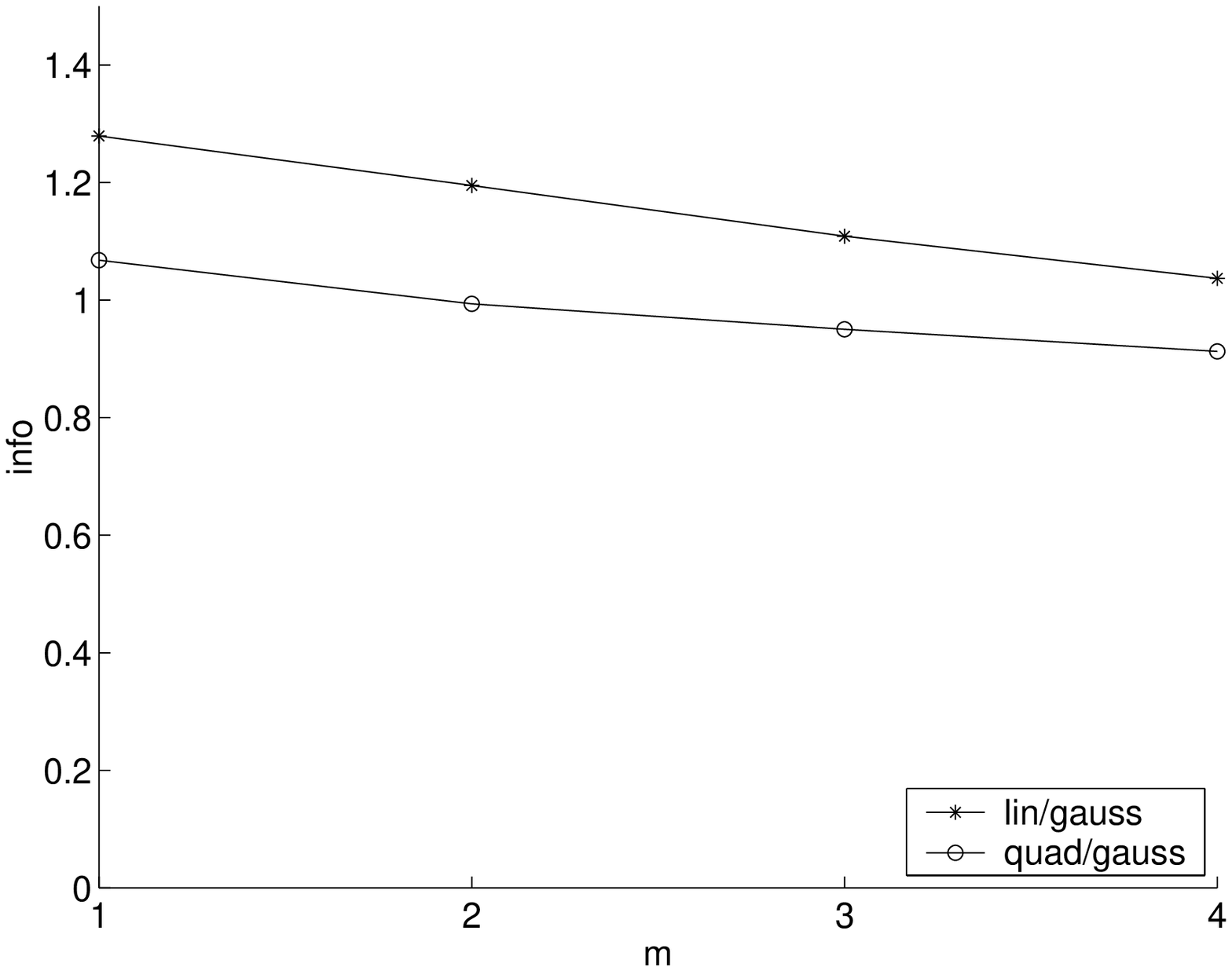,height=6cm,angle=0}}
\caption{Mutual information in linear and quadratic approximation  as in eq.(\protect\ref{final_info4}), for a sample of $10$ cells; $g^2=0.7$; the distribution $\varrho(\varepsilon)$ in eqs.(\protect\ref{lambda1}), (\protect\ref{lambda2}) is just  equal to $1/3$ for each of the three allowed a $\varepsilon$ values of  $0$,$1/2$ and $1$.
 (a)Dependence on the number of movement types $p$. Dotted lines are for the asymptotes, eq.(\protect\ref{largep}); m=1. (b)Dependence on the power $m$ of the cosine in eq.(\protect\ref{tuning2}); p=4.}
\label{res_p_g}
\end{figure}

Fig.\ref{res_vary_g} shows the linear and the quadratic approximations of 
eq.(\ref{final_info4}) for different values of the expansion parameter 
$g^2=\left(\eta^0/2\sigma\right)^2$.
It is easy to see that, for very small values of $g^2$, linear and quadratic approximations roughly coincide, while when $g^2>0.8$ the quadratic approximation begins to fail and one should add higher orders in perturbation theory.

Fig.\ref{res_p_g}(a) shows the dependence of the mutual information on the number of types $p$. The dependence  on $p$ is 
weak for the linear approximation and somewhat stronger for the quadratic one. In both cases an increase in the number of
discrete correlates $p$ produces an increase in the mutual information. The distance between the linear and the quadratic 
approximation remains asymptotically finite (for $p\to\infty$), contrary to what happens in the case of discrete stimuli alone \cite{pap35}.

Fig.\ref{res_p_g}(b) shows the dependence of the mutual information on the width of the directional tuning (see fig.\ref{tuning_2},(b)).
Since we are considering the case when the noise $\sigma$ is large, a very narrow tuning in $\vartheta$ corresponds to a 
larger overlap in the conditional probabilities $p(\eta|\vartheta,s)$, for most angles $\vartheta$. A consequence is that, 
especially in the linear approximation, the mutual information is a (slowly) decreasing function of $m$.

\subsection*{The limit of large $N$}

We consider now the case when the number of neurons is large.
Since we deal with an infinite number of stimuli the mutual information is unbounded. 
Thus we expect that when the number of neurons becomes large and the noise is finite the mutual information tends asymptotically to infinity.

In order to study this limit we discretize the $\{\vartheta\}$ space into a finite set of 
$M=2\pi/\Delta \vartheta$ angles $\vartheta_1..\vartheta_M$,  and then we take the limit $\Delta\vartheta\rightarrow 0$.

The entropy of the responses $H(\{\eta_i\})$ can be written

\begin{equation}
H(\{\eta_i\})=\left\langle\sum_{s=1}^p\sum_{k=1}^M\int \prod_i d\eta_i 
P(\vartheta_k,s) P(\{\eta_i\}|\vartheta_k,s)\log_2\left[
\sum_{s^\prime=1}^p\sum_{k^\prime=1}^M P(s^\prime,\vartheta_{k^\prime})
P(\{\eta_i\}|\vartheta_{k^\prime},s^\prime)\right]
\right\rangle_{\varepsilon,\vartheta^0}
\label{outent2}
\end{equation}
where in analogy with eq.(\ref{dist}) we define

\begin{equation}
P(\{\eta_i\}|\vartheta_k,s)=\prod_{i=1}^N \frac{1}{\sqrt{2\pi\sigma^2}}
exp-\left[\left(\eta_i-\varepsilon^i_s\bar{\eta}_i(\vartheta_k)-(1-\varepsilon_s^i)
\eta_i^f\right)^2/2\sigma^2\right];
\label{dist2}
\end{equation}
and we discretize the average across the directional selectivities $\{\vartheta^0_i\}$ as well

\begin{equation}
\int d\vartheta^0 \varrho(\vartheta^0)\longrightarrow\sum_{k=1}^M\varrho(\vartheta^0_k).
\end{equation}

This situation corresponds to the case when each neuron can discriminate 
across different angles  $\vartheta_1..\vartheta_M$ with a resolution $\Delta \vartheta$.

The calculation can be carried out introducing replicas as in the previous case. One gets

\begin{equation}
H(\{\eta_i\})=-\frac{1}{\ln2}\lim_{n\rightarrow 0}\frac{1}{n}
\left (\sum_{s_1..s_{n+1}=1}^p\sum_{k_1...k_{n+1}=1}^M
\frac{(n+1)^{-\frac{N}{2}}}{(M p)^{n+1}}\frac{1}{(\sqrt{2\pi\sigma^2})^{nN}}
\left\langle exp(-R)^N\right\rangle_{\varepsilon,\vartheta^0}
-1\right )
\label{outent_discr}
\end{equation}
where
\begin{equation}
R=\left[\sum_{l,m}\left (
\varepsilon_{s_l}(\bar{\eta}(\vartheta_{k_l})-\eta^f)-
\varepsilon_{s_m}(\bar{\eta}(\vartheta_{k_m})-\eta^f)\right )^2/
(4\sigma^2 (n+1))\right];
\label{exp_discr}
\end{equation}
and we have assumed symmetry across neurons in the quenched randomness and in the parameters characterizing the conditional distribution $P(\eta_i|s,\vartheta_k)$.

Now we take the limit $N\rightarrow\infty$.
As it is evident from eq.(\ref{exp_discr}), $exp(-R)\leq 1$ and $exp(-R)=1$ when $s_l=s_m$ and $k_m=k_l$
for each pair of indexes ($m$,$l$). 
Thus when $N\rightarrow\infty$ the only terms which survive in the sum on replicas are the ones with $s_1=s_2..=s_{n+1}$ and $k_1=k_2..=k_{n+1}$. Since we have $p$ stimuli $s$ and $M$ stimuli $\vartheta_k$ the total number of terms is $Mp$.
Substituting this value in the sums over replicas in eq.(\ref{outent_discr}) and putting  
$exp(-R)=1$ one obtains an expression for the entropy of the responses $H(\{\eta_i\})$, which summed to the equivocation as in eq.(\ref{final_eq}) gives the final result for the mutual information:

\begin{equation}
I(\{\eta_i\},\vartheta\otimes s)=\log_2(p)+\log_2(M);
\end{equation}

Now we remember that $M=2\pi/\Delta\vartheta$. Taking the limit to continuous angles,
$\Delta\vartheta\rightarrow 0$, it is easy to see that asymptotically the mutual information tends logarithmically to infinity.

\section*{Beyond the gaussian assumption: the {\it tg} model}

So far we have considered the case where the rate distribution for each neuron is normal.
This assumption implies that negative rates have a non zero probability to occur; the more the average rate is small and close to zero, 
the more this probability becomes large. 
The bias introduced by the inclusion of negative rates in the space of possible states might be even more serious since we 
have considered the limit of large noise, where the tail of the distribution in the domain $\eta<0$ acquires a significant weight.
 
Cutting the distribution at zero is not enough to assign the proper weight to under-threshold activity: each time the summation 
of the inputs coming from other units is lower than threshold the neuron remains silent, and this occurs with a well defined probability.

A natural choice for the rate distribution $P(\eta_i|\vartheta,s)$ is a  thresholded gaussian plus a $\delta$ peak in zero ({\it tg} model):

\begin{equation}
P(\eta_i|\vartheta,s)=\frac{1}{\sqrt{2\pi\sigma^2}}
exp-\left[\left(\eta_i-\tilde{\eta}_i(\vartheta,s)
\right)^2/2\sigma^2\right]\Theta(\eta_i)+2(1-\erf(\tilde{\eta}_i(\vartheta,s)/\sigma)\delta(\eta_i)\Theta(-\eta_i)
\label{dist_corr}
\end{equation}
where $\Theta(x)$ is the Heaviside step function,  $\tilde{\eta}_i(\vartheta,s)$ is the same as defined in eq.(\ref{tuning_tot}) and $\erf(x)$ is the error function:

\begin{equation}
\erf(x)=\frac{1}{\sqrt{2\pi}}\int_{-\infty}^{x} dt e^{-t^2/2}.
\label{erf}
\end{equation}

The factor multiplying the $\delta$ function ensures a correct normalization and it assigns the proper weight to the peak in zero, which is larger the more the average rate $\tilde{\eta}_i(\vartheta,s)$ is close to zero.
A similar distribution has already been considered in networks of threshold linear neurons \cite{tre+90,tre+90b}.

The analytical evaluation of the mutual information is obviously more difficult than in the case of  the simple gaussian, because of the presence of the error function, which cannot be integrated exactly.
Nonetheless in the limit of large $\sigma$ it is possible to evaluate  both the linear 
and the quadratic approximation in $N$ and thus quantify the impact of the correction with respect to the gaussian case, eq.(\ref{final_info4}).

\subsection*{The limit of large $\sigma$ for the {\it tg} model: the equivocation}

We remind the expression of the equivocation, eq.(\ref{equiv}):

\begin{equation}
\left\langle H(\{\eta_i\}|\vartheta,s)\right\rangle_{\vartheta,s}=
\left\langle\sum_{s=1}^p\int \!d\vartheta\! \int \prod_i d\eta_i 
P(\vartheta,s) P(\{\eta_i\}|\vartheta,s)\log_2
P(\{\eta_i\}|\vartheta,s)
\right\rangle_{\varepsilon,\vartheta^0}.
\label{equiv1}
\end{equation}

Assuming independence among neurons in the conditional probability 
$P(\{\eta_i\}|\vartheta,s)$, eq.(\ref{equiv1}) can be written 

\begin{equation}
\left\langle H(\{\eta_i\}|\vartheta,s)\right\rangle_{\vartheta,s}=
\frac{N}{\ln 2}\sum_{s=1}^p\int \!d\vartheta\! P(\vartheta,s)\left\langle \int d\eta P(\eta|\vartheta,s)\ln
P(\eta|\vartheta,s)
\right\rangle_{\varepsilon,\vartheta^0}.
\label{equiv2}
\end{equation}

In the specific case of the distribution (\ref{dist_corr}) it is easy to show that
\begin{eqnarray}
&&\!\!\!\left\langle \int d\eta P(\eta|\vartheta,s)\ln
P(\eta|\vartheta,s)
\right\rangle_{\varepsilon,\vartheta^0}\!\!=\nonumber\\
&&\!\frac{1}{\sqrt{2\pi}}\left\langle\frac{\tilde{\eta}(\vartheta,s)}{2\sigma}e^{-\left[\tilde{\eta}(\vartheta,s)\right]^2/2\sigma^2}\right\rangle_{\varepsilon,\vartheta^0}\!\!\!-\!\frac{1}{2}\left(1+\ln(2\pi\sigma^2)\right)\!\left\langle\erf(\tilde{\eta}(\vartheta,s)/\sigma)\right\rangle_{\varepsilon,\vartheta^0}\nonumber\\
&&+\left\langle
\int_0^\infty d\eta \delta(\eta)\left(1-\erf(\tilde{\eta}(\vartheta,s)/\sigma)\right)\left[\ln\delta(\eta)+\ln2+
\ln\left(1-\erf(\tilde{\eta}(\vartheta,s)/\sigma)\right)\right]\right\rangle_{\varepsilon,\vartheta^0}.
\end{eqnarray}

To proceed with the calculation we have to be careful with the integration of the delta 
function. In fact it is easy to show that the integration of the product 
$\delta(x)\ln\delta(x)$ yields a logarithmic divergence.
Since the mutual information must remain finite with a finite number of neurons $N$, we 
expect this divergence to cancel exactly with an analogous term in the rate entropy and, 
in fact, in the next section, we will show that this is the case.
For the moment we use the equality

\begin{equation}
\int_{-\infty}^{+\infty} dx \delta(x) F(x)=\lim_{\epsilon\rightarrow 0}\int_{-\epsilon/2}^{\epsilon/2} dx \frac{1}{\epsilon} F(x).
\label{delta}
\end{equation}

Assuming as usual that the quenched disorder is identically distributed across neurons 
and stimuli and that $\tilde{\eta}(\vartheta,s)$ 
is like in eq.(\ref{dist}) we can write:

\begin{eqnarray}
&&\left\langle H(\{\eta_i\}|\vartheta,s)\right\rangle_{\vartheta,s}=
\frac{N}{2\ln2}\left\{\left(1+\ln(2\pi\sigma^2)\right)
\left\langle\erf(\tilde{\eta}(\vartheta,s)/\sigma)\right\rangle_{\varepsilon,\vartheta^0}
-\left\langle\frac{\tilde{\eta}(\vartheta,s)}{\sqrt{2\pi}\sigma}
e^{-\left[\tilde{\eta}(\vartheta,s)\right]^2/2\sigma^2}
\right\rangle_{\varepsilon,\vartheta^0}\right.\nonumber\\
&&\left.+2\left\langle\left[1-\erf(\tilde{\eta}(\vartheta,s)/\sigma)\right]\right\rangle_{\varepsilon,\vartheta^0}\ln\frac{\epsilon}{2}-
2\left\langle\left[1-\erf(\tilde{\eta}(\vartheta,s)/\sigma)\right]
\ln\left[1-\erf(\tilde{\eta}(\vartheta,s)/\sigma)\right]
\right\rangle_{\varepsilon,\vartheta^0}\right\}.
\label{eq_need_appr}
\end{eqnarray}

The average across quenched disorder cannot be performed if we do not resort 
to some approximation. Since we have already focused on the limit of large $\sigma$ 
it is natural to consider an expansion of the error function in eq.
(\ref{erf}) for a small value of its argument:

\begin{equation}
\erf(x)\simeq \frac{1}{2}+\frac{1}{\sqrt{2\pi}}x+o(x^2);
\label{erf_appr}
\end{equation}

Approximating all the error functions in eq.(\ref{eq_need_appr}) we obtain:

\begin{eqnarray}
&&\left\langle H(\{\eta_i\}|\vartheta,s)\right\rangle_{\vartheta,s}=
\frac{N}{2\ln2}\left\{\frac{1}{2}\left(1+\ln(2\pi\sigma^2)\right)+\ln(\epsilon)\right.\nonumber\\
&&\left.+\left\langle\frac{\tilde{\eta}(\vartheta,s)}{\sqrt{2\pi}\sigma}
\right\rangle_{\varepsilon,\vartheta^0}\left(2+\ln(2\pi\sigma^2)-2\ln\epsilon\right)-
\frac{1}{\pi}\left\langle\frac{\tilde{\eta}(\vartheta,s)^2}{\sigma^2}
\right\rangle_{\varepsilon,\vartheta^0}\right\}+o(1/\sigma^2)
\label{equiv_corr}
\end{eqnarray}
where in line with the approximation used in the case of the simple gaussian distribution we have omitted terms of order
$N/\sigma^k$ with $k>2$.

\subsection*{Evaluation of the mutual information}
We reconsider eq.(\ref{outent}).
Using replicas and assuming that the quenched randomness is identically distributed across neurons  we obtain

\begin{equation}
H(\{\eta_i\})=-\frac{1}{\ln2}\lim_{n\rightarrow 0}\frac{1}{n}
\left( \sum_{s_1..s_{n+1}=1}^p\int \!d\vartheta_1..d\vartheta_{n+1}
 \frac{1}{(2\pi p)^{n+1}}
\left\langle\int d\eta \prod_{k=1}^{n+1}
P(\eta|\vartheta_k,s_k)\right\rangle_{\varepsilon,\vartheta^0}^N
-1\right),
\end{equation}
where $P(\eta|\vartheta,s)$ is given in eq.(\ref{dist_corr}).
Integrating over $d\eta$ yields
\begin{eqnarray}
&&\left\langle\int d\eta \prod_{k=1}^{n+1}
P(\eta|\vartheta_k,s_k)\right\rangle_{\varepsilon,\vartheta^0}^N=
\left\langle(n+1)^{-\frac{1}{2}}\frac{1}{(\sqrt{2\pi\sigma^2})^{n}}
exp(-R)\erf\left(\frac{1}{\sqrt{n+1}}\sum_k\tilde{\eta}(\vartheta_k,s_k)/\sigma\right)\right.\nonumber\\
&&\left.+\left(\frac{2}{\epsilon}\right)^n\prod_k\left[1-\erf\left(\frac{\tilde{\eta}(\vartheta_k,s_k)}{\sigma}\right)
\right]\right\rangle_{\varepsilon,\vartheta^0}^N
\end{eqnarray}
where we have used eq.(\ref{delta}) to integrate the $\delta$ function in eq.(\ref{dist_corr}), and the expression of  $R$ is like  in eq.(\ref{gen_R}).
Using the approximation (\ref{erf_appr}) for the error function and considering the expansion for small $n$
\begin{equation}
a^n\simeq 1+n\ln a
\end{equation}
we obtain
\begin{equation}
H(\{\eta_i\})=-\frac{1}{\ln2}\lim_{n\rightarrow 0}\frac{1}{n}
\left(\sum_{s_1..s_{n+1}=1}^p\int \!d\vartheta_1..d\vartheta_{n+1}
 \frac{1}{(2\pi p)^{n+1}}\left(1-nC-\sum_{k,l\neq k}G_{kl}\right)^N -1\right);
\end{equation}
where
\begin{equation}
C=\frac{1}{2}\left[\frac{1}{2}\left(1+\ln(2\pi\sigma^2)\right)+\ln(\epsilon)+\sum_k\left\langle\frac{\tilde{\eta}(\vartheta_k,s_k)}{\sqrt{2\pi}\sigma}
\right\rangle_{\varepsilon,\vartheta^0}\left(2+\ln(2\pi\sigma^2)-2\ln\epsilon\right)\right];
\label{C}
\end{equation}

\begin{equation}
G_{kl}=\frac{1}{8(n+1)\sigma^2}\left[\left\langle\left(\tilde{\eta}(\vartheta_k,s_k)-\tilde{\eta}(\vartheta_l,s_l)\right)^2\right\rangle_{\varepsilon,\vartheta^0}-\frac{1}{2\pi}\left\langle\tilde{\eta}(\vartheta_k,s_k)\tilde{\eta}(\vartheta_l,s_l)\right\rangle_{\varepsilon,\vartheta^0}\right].
\label{gkl}
\end{equation}

It is simple to verify that $C$ remains finite when $n$ goes to zero.
Now we expand in powers of $N$ up to the second order:

\begin{equation}
\left(1-nC-\sum_{k,l\neq k}G_{kl}\right)^N\simeq 1-N\left(nC+\sum_{k,l\neq k}G_{kl}\right)+\frac{N^2}{2}\sum_{k,l\neq k}\sum_{\varrho,\mu\neq\varrho}G_{kl}G_{\varrho\mu};
\end{equation}
where we have omitted terms that are $o(n)$ when $n\rightarrow 0$.
This quantity has to be summed over continuous and discrete replicas, after having explicitly performed 
the average across quenched disorder in eqs.(\ref{C}) and (\ref{gkl}).
It is easy to show that $C$ cancels exactly with analogous terms in the equivocation, eq.(\ref{equiv_corr}).
The evaluation of the linear and quadratic term in the mutual information can be performed with a similar technique 
to the one used in the case of the gaussian distribution and it involves averaging terms with $2-$ and $4-$replica interactions. 

The final expression for the mutual information in the linear and quadratic approximation reads

\begin{eqnarray}
&&I(\{\eta_i\},\vartheta\otimes s)\simeq\nonumber\\
&&\frac{1}{\ln2}\left\{\frac{N(\eta^0)^2}{4\sigma^2}\left(\frac{1}{2}+\frac{1}{\pi}\right)
\left[\frac{p-1}{p}2\left(\alpha-A_1\right)^2\lambda_1+2\left(A_2-(A_1)^2\right)\lambda_2\right]-\frac{N^2(\eta^0)^4}{2(4\sigma^2)^2}\left(\frac{1}{2}+\frac{1}{\pi}\right)^2\right.\\
&&\left.\left\{\frac{p-1}{p^2}2\left(2\left(\alpha-A_1\right)^2\lambda_1\right)^2+\left[\frac{p-1}{p}\left(\lambda_1-\lambda_2\right)^2+\frac{(\lambda_2)^2}{p}\right]\left[\left(\frac{1}{2^{2m-1}}\right)^4\sum_{\nu=0}^{m-1}\left[\left(\begin{array}{c}2m\\\nu\end{array}\right)\right]^4\right]\right\}\right\}.\nonumber
\label{final_info4_corr}
\end{eqnarray}

Comparing eq.(\ref{final_info4}) and (\ref{final_info4_corr}) it is evident that  modifying the gaussian model into the more realistic {\it tg} model has no effect on the analytical expression of the mutual information, except for a renormalization of the expansion parameter $g^2=(\eta^0/2\sigma)^2$:

\begin{equation}
g^\prime=g\sqrt{\frac{1}{2}+\frac{1}{\pi}}.
\end{equation}

\begin{figure}
\centerline{
\psfig{figure=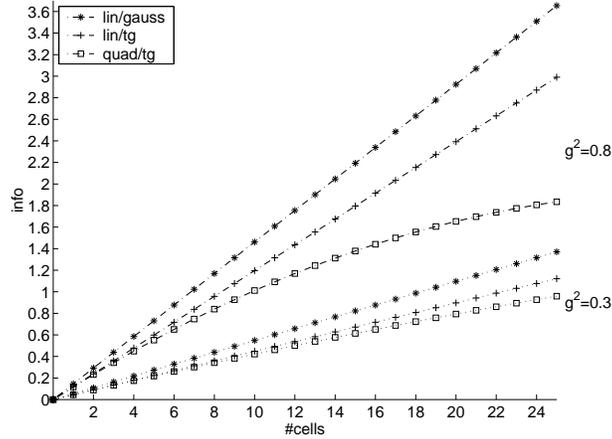,height=6cm,angle=0}}
\caption{Information rise as in eqs.(\protect\ref{final_info4}),
(\protect\ref{final_info4_corr}) for different values of the expansion parameter  $g^2=\left(\eta^0/2\sigma\right)^2$; $m=1$; $p=4$; the distribution 
$\varrho(\varepsilon)$ in eqs.(\protect\ref{lambda1}), (\protect\ref{lambda2}) 
is just equal to $1/3$ for each of the 3 allowed $\varepsilon$ values of $0$,$1/2$ and $1$.}
\label{res_vary}
\end{figure}

Fig.\ref{res_vary} shows the effect of the renormalization for different values of  
$g^2$.
The mutual information is lower in the {\it tg} model than in the gaussian 
approximation, as expected.

We have explored whether the two models can fit the information rise estimated 
from real data.
Since the analytical expression of the mutual information is the same in both cases 
the fit does not change between the two models. 

\begin{figure}
\centerline{
\psfig{figure=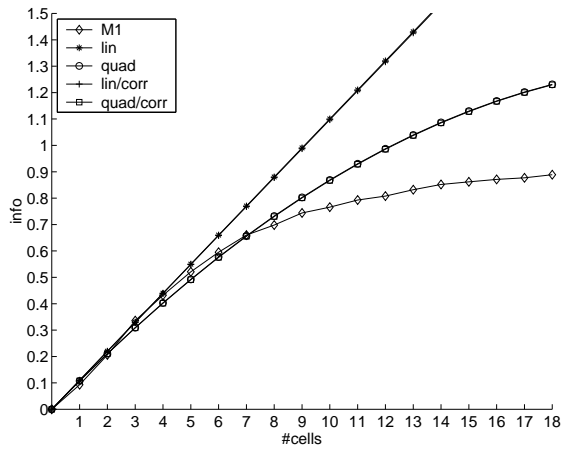,height=6cm,angle=0}}
\caption{Comparison between the theoretical curves, eq.(\protect\ref{final_info4}), (\protect\ref{final_info4_corr}) and the information estimated from a sample of cells 
recorded in the right primary motor cortex \protect\cite{vale}; m=1;p=2; 
the distribution $\varrho(\varepsilon)$ in eqs.(\protect\ref{lambda1}),
(\protect\ref{lambda2}) is just equal to $1/3$ for each of the three allowed  $\varepsilon$ of $0$,$1/2$,$1$; the values of $g^2=\left(\eta^0/2\sigma\right)^2$ used for the 
fit are $0.64$ for eq.(\protect\ref{final_info4}) and $0.78$ for 
eq.(\protect\ref{final_info4_corr})}
\label{res_fit}
\end{figure}

Fig.\ref{res_fit} shows the comparison between the information estimated from a sample 
of cells recorded in the right primary motor cortex \cite{vale} and the prediction 
given by either of the two theoretical distributions.
In the limit of large $\sigma$ the gaussian model fails to provide a good fit, but we 
can conclude that this failure 
is not due to the inclusion of negative rates in the distribution. 
   
\section*{Information loss in averaging activity distributions}
Fig.\ref{tuning_data} suggests that the directional tuning of real cells is
modulated, albeit moderately, by the type of movement.
In fact the analysis of real data has proved that the coding of the direction is not 
unique, but it is specific to the complex correlate which is being considered \cite{vale} (here, which arm moves).

More in general, distinct features characterizing a complex stimulus are not expected to 
be coded independently of one another. 
This raises the question of how central representations of external correlates are 
constructed and which are the basic featural components of these representations.
Of course the categorization of natural stimuli is arbitrary and the more accurate a 
description is provided, the higher the dimensionality of the stimulus set. 
Since an infinite number of different descriptors could be chosen to characterize a 
stimulus, any (finite) categorization has the effect of emphasizing some {\it relevant} features 
and averaging out other {\it irrelevant} features.
An obvious consequence is that we end up, even involuntarily, evaluating how some 
features are coded {\it on average}, with respect to the dimension we have chosen, explicitly or implicitly, to neglect.

Thus, with correlates which have one continuous and one discrete dimension, one might 
wonder which are the relationships among the information carried about the total number 
of continuous+discrete dimensions, the information carried about the continuous dimension, 
disregarding the discrete dimension and, finally, the information carried about the 
continuous dimension,
if a single value of the discrete dimension had been fixed when recording neural activity.
In other words, suppose that we investigate how the direction is coded {\it on average}
across different types of movement.
This corresponds to averaging the full distribution $P(\{\eta_i\}|\vartheta,s)$ on $s$:

\begin{equation}
P(\{\eta_i\}|\vartheta,s)\rightarrow \sum_s P(s) P(\{\eta_i\}|\vartheta,s)=P_s(\{\eta_i\}|\vartheta).
\end{equation}

The resulting expression of the mutual information is:

\begin{equation}
I(\{\eta_i\},\langle\vartheta\rangle_s)=
\left\langle\int \!d\vartheta\! \int \prod_i d\eta_iP(\vartheta) 
P_s(\{\eta_i\}|\vartheta)\log_2\frac{P_s(\{\eta_i\}|\vartheta)}{P(\{\eta_i\})}
\right\rangle_{\varepsilon,\vartheta^0}.
\label{info_avg}
\end{equation}

The analytical evaluation is very similar to the cases 
already discussed.

As usual, the mutual information can be expressed as the difference between  
the entropy of the responses $H(\{\eta_i\})$ and the equivocation 
$\left\langle H(\{\eta_i\}|\vartheta)\right\rangle_{\vartheta}$, where 

\begin{equation}
\left\langle H(\{\eta_i\}|\vartheta)\right\rangle_{\vartheta}=
\left\langle\int \!d\vartheta\! \int \prod_i d\eta_i 
P(\vartheta) P_s(\{\eta_i\}|\vartheta)\log_2
P_s(\{\eta_i\}|\vartheta)
\right\rangle_{\varepsilon,\vartheta^0};
\label{equiv_avg}
\end{equation}

\begin{equation}
H(\{\eta_i\})=\left\langle\int \!d\vartheta\! \int \prod_i d\eta_i 
P(\vartheta) P_s(\{\eta_i\}|\vartheta)\log_2\left[
\int \!d\vartheta^\prime P(\vartheta^\prime)
P_s(\{\eta_i\}|\vartheta^\prime)\right]
\right\rangle_{\varepsilon,\vartheta^0}.
\label{outent_avg}
\end{equation}

We focus on the more realistic {\it tg} model, eq.(\ref{dist_corr}); the entropy of the 
responses is obviously independent 
on the chosen categorization of the stimuli; in the limit of large $\sigma$ the 
procedure  is precisely the one followed in the previous section.

The difference with respect to the cases already discussed is in the evaluation 
of the equivocation: since $P_s(\{\eta_i\}|\vartheta)$ is obtained averaging the 
distribution (\ref{dist_corr}) across $s$, we need to introduce discrete replicas 
$s_1..s_n$, as when evaluating the entropy of the responses. Then 
the calculation is straightforward and the basic steps are given in the previous section 
and in the appendix.

Since in one case (for the entropy of the responses) we sum  both over discrete and over 
continuous replicas and in the other case (for the equivocation) we sum only on 
discrete replicas, it is clear that all terms that do not involve 2 or more 
replica interacting cancel out.

More in detail, if the evaluation of the entropy of the responses requires the analytical 
calculation of averages like 
$\left\langle\bar{\eta}(\vartheta_k)\bar{\eta}(\vartheta_l)\right\rangle_{\vartheta^0}$ 
(see the appendix), 
these averages disappear in the evaluation of the equivocation, since replica indexes 
are only for the discrete variable $s$.

The final result for the mutual information up to the quadratic approximation reads:

\begin{eqnarray}
&&I(\{\eta_i\},\left\langle\vartheta\right\rangle_s)\simeq
\frac{1}{\ln2}\left\{\frac{N(\eta^0)^2}{4\sigma^2}
\left(1+\frac{2}{\pi}\right)\left(A_2-(A_1)^2\right)\left[\lambda_2-\frac{p-1}{p}\lambda_1\right]\right .\nonumber\\
&&\left .-\frac{N^2(\eta^0)^4}{2(4\sigma^2)^2}
\left(\frac{1}{2}+\frac{1}{\pi}\right)^2\left\{\frac{p-1}{p^2}
8\left[\left(\alpha-A_1\right)^4-\left(A_2-2\alpha A_1+\alpha^2\right)^2\right]\left(\lambda_1\right)^2\right.\right.\nonumber\\
&&\left.\left.+\left[\frac{p-1}{p}\left(\lambda_1-\lambda_2\right)^2+\frac{(\lambda_2)^2}{p}\right]
\left[\left(\frac{1}{2^{2m-1}}\right)^4\sum_{\nu=0}^{m-1}\left[\left(\begin{array}{c}2m\\\nu\end{array}\right)\right]^4\right]\right\}\right\}.
\label{final_info4_avg}
\end{eqnarray}

\begin{figure}
\hbox{
\mbox{(a)}
\psfig{figure=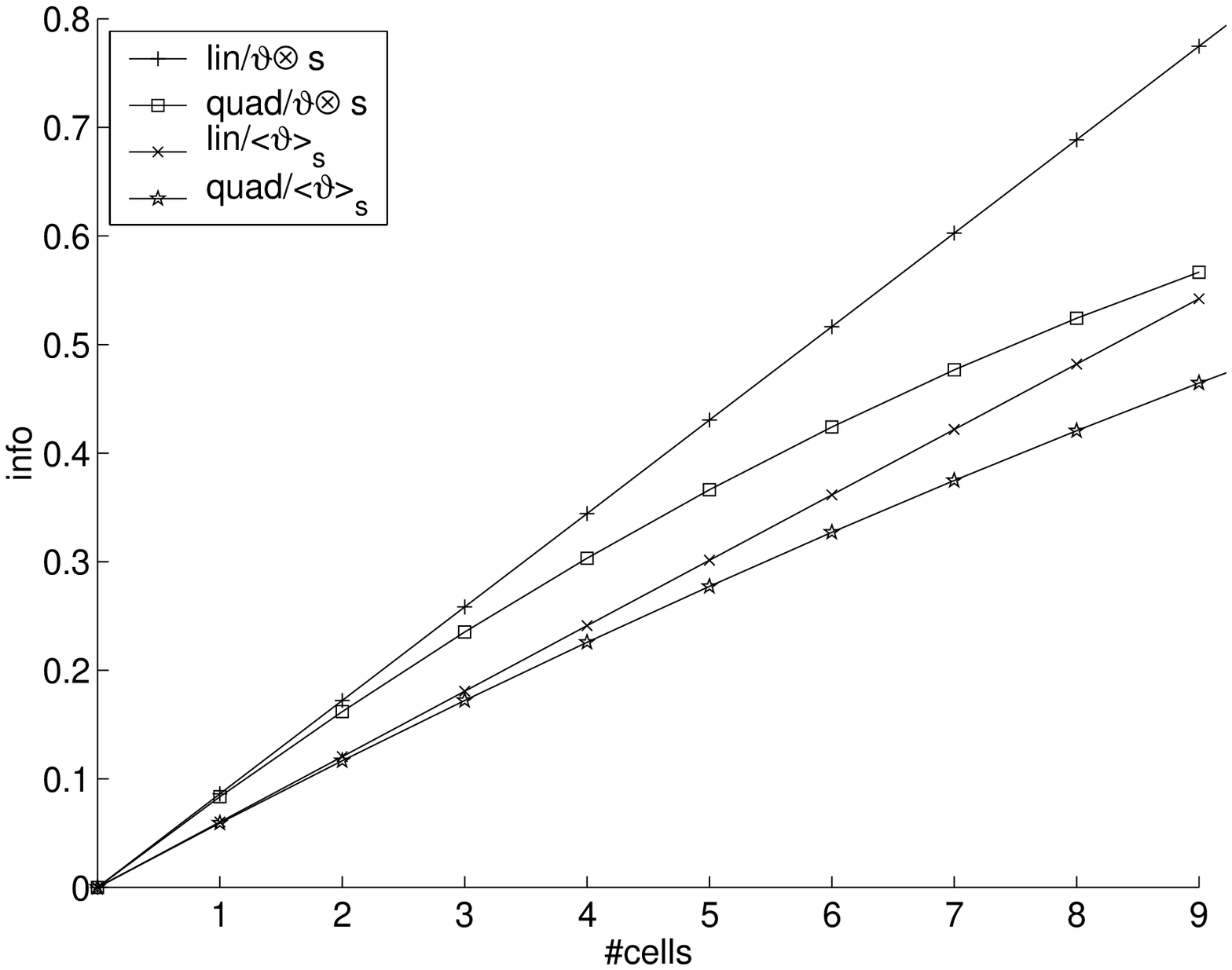,height=6cm,angle=0}
\mbox{(b)}
\psfig{figure=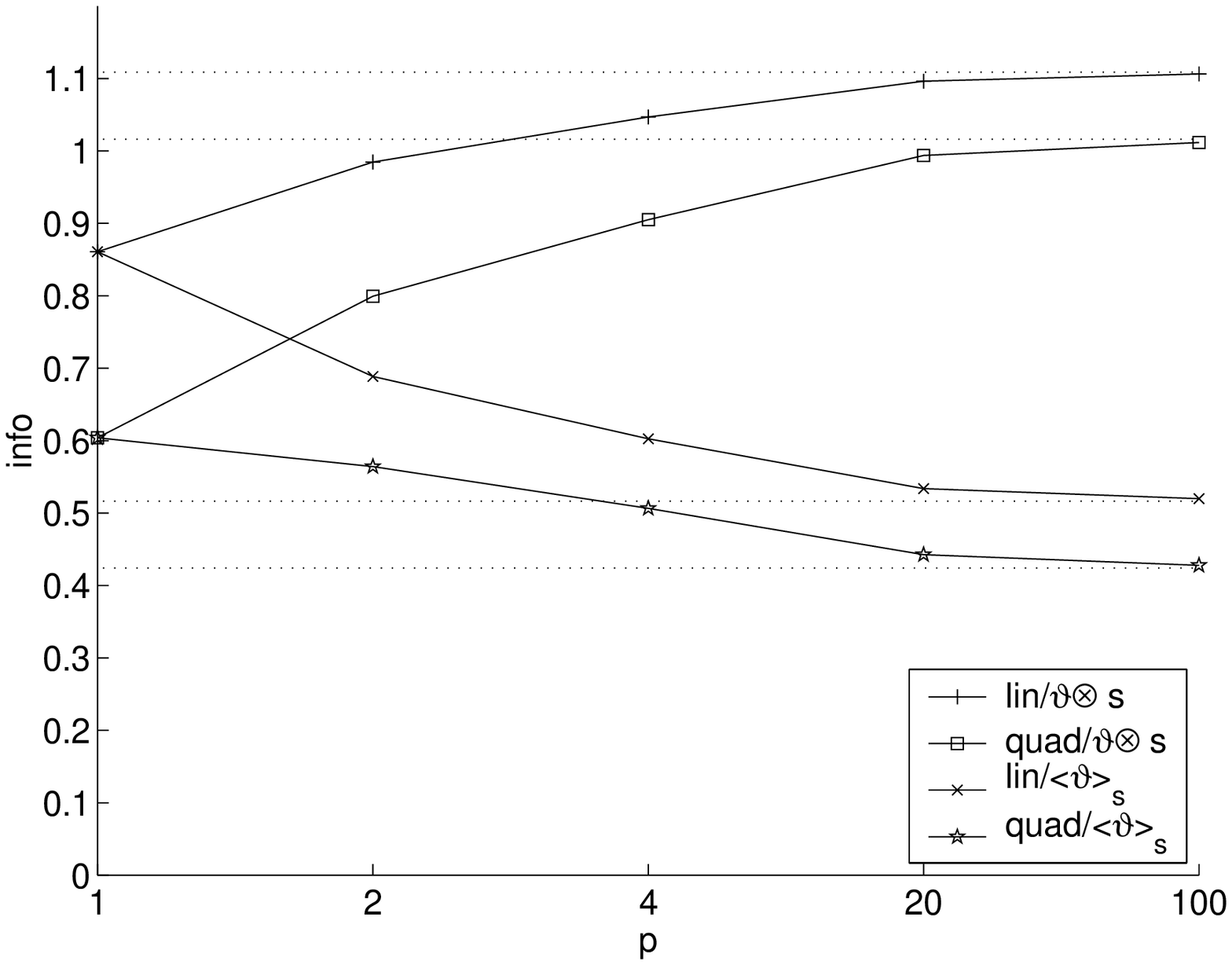,height=6cm,angle=0}}
\caption{Mutual information in the linear and quadratic approximations, as in 
eqs.(\protect\ref{final_info4_corr}), (\protect\ref{final_info4_avg}), for a sample of 10 cells; $g^2=0.7$; $m=1$; 
the distribution $\varrho(\varepsilon)$ 
in eqs.(\protect\ref{lambda1}), (\protect\ref{lambda2}) 
is just equal to $1/3$ for each of the 3 allowed $\varepsilon$ values of $0$,$1/2$ and $1$.
(a) Full curve as a function of the population size; $p=1$ for the {\it full} information 
$I(\{\eta_i\},\vartheta\otimes s)$ and $p=4$ for the {\it averaged } information 
$I(\{\eta_i\},\left\langle\vartheta\right\rangle_s)$. 
(b) Dependence on the number of movement types $p$. 
Dotted lines are for the asymptotes, $p\rightarrow\infty$}
\label{res_p_avg}
\end{figure}

Fig.\ref{res_p_avg}(a) compares the {\it averaged} information, $I(\{\eta_i\},\left\langle\vartheta\right\rangle_s)$, with the {\it full} information $I(\{\eta_i\},\vartheta\otimes s)$, where in the case of $I(\{\eta_i\},\vartheta\otimes s)$  we have put $p=1$. 
The full information calculated with $p=1$ in fact gives the curve one would obtain, on average,
by considering only one movement type (or value of the discrete correlate) at a time.
As one could expect, averaging the distribution across the discrete correlates 
results in an information loss. Moreover, the the {\it full} information  with $p=4$ movement types
is obviously above the {\it specific} one, obtained by setting $p=1$ (compare fig.\ref{res_vary} and \ref{res_p_avg}(a)). 

Fig.\ref{res_p_avg}(b) shows the dependence of the full and averaged information on the number $p$ of discrete correlates.
Contrary to the {\it full} information, the {\it averaged} information 
decreases monotonically with $p$, both in the linear and in quadratic approximation.

As one would expect, averaging the distribution across a large number $p$ of correlates 
is equivalent to a regularization of the activity distribution, which results in a lower mutual information.

\section*{Discussion}
We have studied a model of the coding of discrete+continuous stimuli by a population of $N$ neurons, 
referring to the specific case of movements categorized according to a direction and a type. 
We have shown that, asymptotically in the limit of large populations of neurons, the mutual information  
tends to infinity logarithmically with the resolution in the continuous dimension. This result aside,
we have focused on the initial rise of the mutual information with the population size,
which may offer a more direct comparison with the analysis of real data.

In the limit of large noise we have therefore derived an analytical formula for such initial rise of the  
information, up to the quadratic approximation. We have examined the dependence on the number of discrete
correlates and on the width of the directional tuning. A comparison with the information estimated from real 
data, in which the linear term is used as a fit coefficient, has shown that the quadratic approximation fails
to capture the deviation from linearity.

We have then considered a more realistic model for the conditional firing distribution, a thresholded gaussian
with a $\delta$-peak. We have shown that this more realistic distribution simply renormalizes the expansion
parameter applicable to the gaussian distribution. Therefore, the discrepancy in the fit to real data
does not originate in the firing rate distribution. There are several possible reasons for this discrepancy:
\begin{itemize}
\item the value of the expansion parameter $g^2=(\eta^0/\sigma)^2$ corresponding to the best fit in fig.(\ref{res_fit}) 
is quite high ($g^2=0.78$). This value is in the range where the quadratic approximation is expected to fail
on its own (see fig.(\ref{res_vary_g}),(\ref{res_vary})). Adding higher orders in perturbation theory might
improve the fit. Moreover, we have neglected terms of order $N/\sigma^k$ with $k>2$, which might be non 
negligible when $g^2$ becomes large and $N$  is not too small. 
\item information estimates from real data are often biased because of poor 
sampling. Several procedures have been proposed to correct the bias \cite{Pan+96b},
but the improvement given by the correction is not precisely quantifiable, and sampling biases cannot be discarded for good.
\item both with the {\it tg} model and in the gaussian approximation, we have assumed 
that neurons fire independently of one another to each movement, but the analysis of 
extracellular recordings has shown that correlations may play a non-negligible role in 
the coding \cite{pan+99,pan+99b}.
\end{itemize}

Finally, we have examined the effect of averaging the distribution across 
the discrete correlates, evaluating the mutual information with respect to the 
continuously varying dimension alone.
As expected, averaging the distribution across $s$ results in an information loss, 
which is more serious the larger the number $p$ of discrete correlates.

Further developments of this work include: the introduction of weak correlations 
in the signal of different neurons and the analysis of the information transfer to 
other stages of processing, given this as the coding scheme in the input layer of the 
network. In the specific case of movement coding these research directions might help model 
how information is transmitted down the motor system, in the planning and 
execution of motor tasks.
 
\appendix
\section{Detailed evaluation of the second order coefficient}

We show here how to evaluate the coefficient of the quadratic term in eq.(\ref{R_appr}), 
that we write again:

\begin{equation}
\sum_{k,l\neq k}\sum_{\varrho,\mu\neq \varrho}\!
\left\langle\left (\varepsilon_{s_k}(\bar{\eta}(\vartheta_k)\!-\!\eta^f)\!-\!
\varepsilon_{s_l}(\bar{\eta}(\vartheta_l)\!-\!\eta^f)\right )^2\right\rangle\!_{\varepsilon,\vartheta^ 0}\left\langle
\left (\varepsilon_{s_\varrho}(\bar{\eta}(\vartheta_\varrho)\!-\!\eta^f)\!-\!
\varepsilon_{s_\mu}(\bar{\eta}(\vartheta_\mu)\!-\!\eta^f)\right )^2
\right\rangle_{\varepsilon,\vartheta^0}.
\end{equation}

This quantity has to be integrated over continuous and discrete replicas.
First we perform the average across the quenched variable $\vartheta_0$.
Expanding the products it is easy to see that the quantities to be averaged are all 
like $\left\langle\bar{\eta}(\vartheta_k)\right\rangle_{\vartheta^0}$ 
and $\left\langle\bar{\eta}^2(\vartheta_k)\right\rangle_{\vartheta^0}$, 
that we have already calculated in eq.(\ref{A_1}),(\ref{A_2}). 
Another average that we need is 
$\left\langle\bar{\eta}(\vartheta_k)\bar{\eta}(\vartheta_l)\right\rangle_{\vartheta^0}$, with $k\neq l$:

\begin{equation}
\left\langle\bar{\eta}(\vartheta_k)\bar{\eta}(\vartheta_l)\right\rangle_{\vartheta^0}=
\left(\eta^0\right)^2\left\{\left(\frac{1}{2^{2m}}\right)^2\left[\left(\begin{array}{c}2m\\m\end{array}\right)\right]^2+\frac{1}{2}\left(\frac{1}{2^{2m-1}}\right)^2\sum_{\nu=0}^{m-1}\left[\left(\begin{array}{c}2m\\\nu\end{array}\right)\right]^2\cos\left\{\left(m-\nu\right)\left(\vartheta_k-\vartheta_l\right)\right\}\right\}.
\label{A_3}
\end{equation}

Thus only terms like $\left\langle\bar{\eta}(\vartheta_k)\bar{\eta}(\vartheta_l)\right\rangle_{\vartheta^0}$ still depend on continuous replicas. 

Since terms like $\cos(m-\nu)(\vartheta_k-\vartheta_l)$, with $k\neq l$ are zero when integrated on $\vartheta_k$,$\vartheta_l$ the only term which requires a careful evaluation in performing the integration on continuous replicas is 
the product $\left\langle\bar{\eta}(\vartheta_k)\bar{\eta}(\vartheta_l)\right\rangle_{\vartheta^0}\left\langle\bar{\eta}(\vartheta_{\varrho})\bar{\eta}(\vartheta_\nu)\right\rangle_{\vartheta^0}$.

After integration on continuous replicas this term yields

$$
\left(\eta^0\right)^4\left\{\left(\frac{1}{2^{2m}}\right)^4\left[\left(\begin{array}{c}2m\\m\end{array}\right)\right]^4+\frac{1}{4}\left(\frac{1}{2^{2m-1}}\right)^4\sum_{\nu=0}^{m-1}\left[\left(\begin{array}{c}2m\\\nu\end{array}\right)\right]^4\left(\delta_{k\varrho}\delta_{\mu l}+\delta_{k\mu}\delta_{\varrho l}\right)\right\}.
$$

Rearranging all terms one has, finally

\begin{eqnarray}
&&\sum_{k,l\neq k}\sum_{\varrho,\mu\neq\varrho}\!\left [\left(\alpha-A_1\right)^2\left\langle(\varepsilon_{s_k}\!-\!\varepsilon_{s_l})^2\right\rangle_{\varepsilon}+\left(A_2\!-\!(A_1)^2\right)\left\langle\varepsilon_{s_k}^2\!+\!\varepsilon_{s_l}^2\right\rangle_{\varepsilon}\right]x\nonumber\\
&&x\left[\left(\alpha-A_1\right)^2\left\langle(\varepsilon_{s_{\varrho}}\!-\!\varepsilon_{s_{\mu}})^2\right\rangle_{\varepsilon}+\left(A_2\!-\!(A_1)^2\right)\left\langle\varepsilon_{s_{\varrho}}^2\!+\!\varepsilon_{s_{\mu}}^2\right\rangle_{\varepsilon}\right]\nonumber\\
&&+\left\langle\varepsilon_{s_k}\varepsilon_{s_l}\right\rangle_{\varepsilon}\left\langle\varepsilon_{s_{\varrho}}\varepsilon_{s_{\mu}}\right\rangle_{\varepsilon}\left(\frac{1}{2^{2m-1}}\right)^4\sum_{\nu=0}^{m-1}\left[\left(\begin{array}{c}2m\\\nu\end{array}\right)\right]^4\left(\delta_{k\varrho}\delta_{\mu l}+\delta_{k\mu}\delta_{\varrho l}\right)
\label{mid_sum}
\end{eqnarray}
where $A_1$ and $A_2$ are defined in eq.(\ref{A_1}) and (\ref{A_2}).
To correctly perform the summation across the discrete replicas and to keep only terms 
of order $n$ in the limit 
$n\rightarrow 0$, we consider separately the different contributions to the sum:

\begin{equation}
\sum_{k,l\neq k}\sum_{\varrho,\mu\neq\varrho}\Rightarrow
\sum_{k\neq l\neq\varrho\neq \mu}+\sum_{\varrho,k\neq l\neq \mu}\!\left(\delta_{\varrho k}+
\delta_{\varrho l}\right)+\!\sum_{\mu,k\neq l\neq\varrho}\!\left(\delta_{\mu k}+
\delta_{\mu l}\right)
+\!\sum_{\mu,\varrho, k\neq l}\!\left(\delta_{\mu k}\delta_{\varrho l}+\delta_{\varrho k}\delta_{\mu l}\right).
\label{sum_sep}
\end{equation}

Since in eq.(\ref{mid_sum}) terms like $\left\langle(\varepsilon_{s_k}-\varepsilon_{s_l})^2\right\rangle_{\varepsilon}$ are zero if $s_k=s_l$, we have to distinguish between different cases for each term in eq.(\ref{sum_sep}) in performing the summation on discrete replicas: 

\begin{eqnarray}
&&\sum_{s_1}..\sum_{s_{n+1}}\Rightarrow\sum_{s_1}..\sum_{s_k\neq s_l}\sum_{s_{\varrho}\neq s_m}..\sum_{s_{n+1}}+\sum_{s_1}..\sum_{s_k= s_l}\sum_{s_{\varrho}\neq s_m}..\sum_{s_{n+1}}\nonumber\\
&&+\sum_{s_1}..\sum_{s_k\neq s_l}\sum_{s_{\varrho}=s_m}..\sum_{s_{n+1}}+\sum_{s_1}..\sum_{s_k= s_l}\sum_{s_{\varrho}= s_m}..\sum_{s_{n+1}};
\end{eqnarray}

With a bit of combinatorics and keeping only the terms order $n$, which give a finite contribution in the limit $n\rightarrow 0$, we obtain the final result for the second order term

$$
\frac{N^2(\eta^0)^4}{2(4\sigma^2)^2}\left\{\frac{p-1}{p^2}2\left(2 \left(\alpha-A_1\right)^2\lambda_1\right)^2+\left[\frac{p-1}{p}\left(\lambda_1-\lambda_2\right)^2+\frac{\lambda_2^2}{p}\right]\left[\left(\frac{1}{2^{2m-1}}\right)^4\sum_{\nu=0}^{m-1}\left[\left(\begin{array}{c}2m\\\nu\end{array}\right)\right]^4\right]\right\};
$$
where the expressions of $\lambda_1$,$\lambda_2$ are given in eqs.(\ref{lambda1}), (\ref{lambda2}). 
Considering the result obtained at the first order, eq.(\ref{final_info2}), it is easy to derive the expression for the mutual information up to the second order in $N/\sigma^2$:

\begin{eqnarray*}
&&I(\{\eta_i\},\vartheta\otimes s)\simeq\frac{1}{\ln2}\left\{\frac{N(\eta^0)^2}{4\sigma^2}
\left[\frac{p-1}{p}2\left(\alpha-A_1\right)^2\lambda_1+2\left(A_2-(A_1)^2\right)\lambda_2\right]\right .\nonumber\\
&&\left .-\frac{N^2(\eta^0)^4}{2(4\sigma^2)^2}\left\{\frac{p-1}{p^2}2\left(2\left(\alpha-A_1\right)^2\lambda_1\right)^2+\left[\frac{p-1}{p}\left(\lambda_1-\lambda_2\right)^2+\frac{\lambda_2^2}{p}\right]\left[\left(\frac{1}{2^{2m-1}}\right)^4\sum_{\nu=0}^{m-1}\left[\left(\begin{array}{c}2m\\\nu\end{array}\right)\right]^4\right]\right\}\right\}.
\label{final_infoapp}
\end{eqnarray*}

\subsection*{Acknowledgements}
We have enjoyed extensive discussions with In\'es Samengo, Eilon Vaadia, Edmund Rolls and Elka Korutcheva. Partial support from Human Frontier Science Programme grant RG 0110/1998-B.

\end{document}